\documentclass[10pt,conference]{IEEEtran}

\usepackage[dvips]{graphicx}
\usepackage{psfrag}
\usepackage[ieee]{research5}
\newtheorem{theorem}{Theorem}

\newtheorem{corollary}[theorem]{Corollary}

\author{\authorblockN{Stefan M.~Moser}
\authorblockA{Department of Communication Engineering\\
National Chiao Tung University (NCTU)\\
Hsinchu, Taiwan \\
Email: stefan.moser@ieee.org}
}

\title{On the Fading Number of Multiple-Input Single-Output Fading
  Channels with Memory}

\begin{document}

\maketitle

\begin{abstract}
  We derive new upper and lower bounds on the fading number of
  multiple-input single-output (MISO) fading channels of general (not
  necessarily Gaussian) regular law with spatial and temporal memory.
  The fading number is the second term, after the double-logarithmic
  term, of the high signal-to-noise ratio ($\SNR$) expansion of channel
  capacity.
  
  In case of an isotropically distributed fading vector it is proven
  that the upper and lower bound coincide, \emph{i.e.}, the general
  MISO fading number with memory is known precisely.
  
  The upper and lower bounds show that a type of beam-forming is
  asymptotically optimal.
  
\end{abstract}


\section{Introduction}
\label{sec:introduction}

It has been recently shown in \cite{lapidothmoser03_3}, \cite{moser04}
that, whenever the matrix-valued fading process is of finite
differential entropy rate, the capacity of multiple-input
multiple-output (MIMO) fading channels typically grows only
double-logarithmically in the signal-to-noise ratio ($\SNR$).  To
quantify the rates at which this poor power efficiency begins,
\cite{lapidothmoser03_3}, \cite{moser04} introduced the \emph{fading
  number} as the second term in the high-$\SNR$ asymptotic expansion
of channel capacity. Explicit expressions for the fading number were
then given for a number of fading models. For channels with memory,
the fading number of single-input single-output (SISO) fading channels
was derived in \cite{lapidothmoser03_3}, \cite{moser04} and the
single-input multiple-output (SIMO) case was derived in
\cite{lapidothmoser04_2}, \cite{lapidothmoser04_3app}, \cite{moser04}.

The fading number of the multiple-input single-output (MISO) fading
channel has been derived in general only for the memoryless
case \cite{lapidothmoser03_3}, \cite{moser04}:
\begin{equation}
  \label{eq:iidmiso}
  \chi(\trans{\vect{H}}) = \sup_{\|\hvect{x}\|=1} \left\{ \log \pi +
    \E{\log|\trans{\vect{H}} \hvect{x}|^2} -
    h(\trans{\vect{H}}\hvect{x}) \right\}.
\end{equation}
This fading number is achievable by inputs that can be expressed as
the product of a constant unit vector in $\Complex^{\nt}$ and a
circularly symmetric, scalar, complex random variable of the same law
that achieves the memoryless SISO fading number
\cite{lapidothmoser03_3}. Hence, the asymptotic capacity of a MISO
fading channel is achieved by beam-forming where the beam-direction is
chosen not to maximize the $\SNR$, but the fading number.

In \cite{kochlapidoth05_1} and \cite{kochlapidoth05_2} Koch \&
Lapidoth investigate the fading number of MISO fading channels with
memory where the fading is Gaussian. For the case of a mean-$\vect{d}$
Gaussian vector process with memory where $\{\vect{H}_k-\vect{d}\}$ is
spatially independent and identically distributed (IID) and where each
component is a zero-mean unit-variance circularly symmetric complex
Gaussian process, the fading number is shown to be\footnote{Note that
  all results in this paper are in nats.}
\begin{IEEEeqnarray}{rCl}
  \chi_{\textnormal{Gauss, spat.~IID}}(\{\trans{\vect{H}}_k\}) 
  & = & -1 +\log\|\vect{d}\|^2 - \Ei{-\|\vect{d}\|^2} \nonumber \\
  && +\: \log\frac{1}{\eps^2},  
  \label{eq:iidgauss}
\end{IEEEeqnarray}
where $\eps^2$ denotes the prediction error when predicting one of
the components of the fading vector based on the observation of its
past.

Furthermore, Koch \& Lapidoth derive an upper bound to the fading
number for the general Gaussian case, \emph{i.e.},
$\{\vect{H}_k-\vect{d}\}$ is a zero-mean circularly symmetric
stationary ergodic complex Gaussian process with matrix-valued
spectral distribution function $\mat{F}(\cdot)$ and with covariance
matrix $\mat{K}$. Assuming that the prediction error covariance matrix
$\mat{\Sigma}$ is non-singular (regularity assumption) they show that
\begin{equation}
  \label{eq:gauss}
  \chi_{\textnormal{Gauss}}(\{\trans{\vect{H}}_k\}) \le -1 +
  \log d_{\ast}^2 - \Ei{-d_{\ast}^2} + \log
  \frac{\|\mat{K}\|}{\lambda_{\textnormal{min}}}, 
\end{equation}
where 
\begin{equation}
  \label{eq:dgauss}
  d_{\ast} = \max_{\|\hvect{x}\|=1} \frac{\left|
      \E{\trans{\vect{H}}_k}\hvect{x}
    \right|}{\sqrt{\Var{\trans{\vect{H}_k}\hvect{x}}}};
\end{equation}
$\lambda_{\textnormal{min}}$ denotes the smallest eigenvalue of
$\mat{\Sigma}$; and where $\|\cdot\|$ denotes the Euclidean operator
norm of matrices, \emph{i.e.}, the largest singular value.

In this paper we extend these results to general (not necessarily
Gaussian) fading channels. 

The remaining of this paper is structured as follows: after defining
the channel model in detail in the following section, we will present
the main results, \emph{i.e.}, a new upper and lower bound on the
MISO fading number, in Section~\ref{sec:main}. 

We then specialize these results to the case of isotropically
distributed fading processes in Section~\ref{sec:csfading} and to
Gaussian fading in Section~\ref{sec:gaussianfading}. For isotropically
distributed fading we will show that the upper and lower bound
coincide. In the Gaussian case we shall derive the above mentioned
results of Koch \& Lapidoth as special cases of our bounds.



We conclude in Section~\ref{sec:conclusion}.

\section{The Channel Model}
\label{sec:channel-model}

We consider a MISO fading channel whose time-$k$ output
$Y_k\in\Complex$ is given by
\begin{equation}
  \label{eq:channel_model}
  Y_k = \trans{\vect{H}_k}\vect{x}_k + Z_k
\end{equation}
where $\vect{x}_k\in\Complex^{\nt}$ denotes the time-$k$ channel input
vector; where the random vector $\vect{H}_k$ denotes the time-$k$
fading vector; where $\trans{\vect{H}_k}$ denotes the transpose of the
vector $\vect{H}_k$; and where $Z_k$ denotes additive noise. Here
$\Complex$ denotes the complex field, $\Complex^{\nt}$ denotes the
$\nt$-dimensional complex Euclidean space, and $\nt$ is the number of
transmit antennas. We assume that the additive noise is an IID
zero-mean white Gaussian process of variance $\sigma^2>0$.

As for the multi-variate fading process $\{\vect{H}_{k}\}$, we shall
only assume that it is stationary, ergodic, of finite second moment
\begin{equation}
  \label{eq:energy}
  \E{\|\vect{H}_{k}\|^{2}} < \infty,
\end{equation}
and of finite differential entropy rate
\begin{equation}
  \label{eq:entropy}
  h(\{\vect{H}_{k}\}) > -\infty.
\end{equation}

Finally, we assume that the fading process $\{\vect{H}_{k}\}$ and the
additive noise process $\{Z_{k}\}$ are independent and of a
joint law that does not depend on the channel input $\{\vect{x}_{k}\}$. 

As for the input, we consider two different constraints: a peak-power
constraint and an average-power constraint. We use $\Es$ to denote the
maximal allowed instantaneous power in the former case, and to denote
the allowed average power in the latter case. For both cases we set
\begin{equation}
  \SNR \eqdef \frac{\Es}{\sigma^{2}}.  
\end{equation}

The capacity $\const{C}(\SNR)$ of the channel
\eqref{eq:channel_model} is given by
\begin{equation}
  \const{C}(\SNR) = \lim_{n \rightarrow \infty} \frac{1}{n} \sup 
  I\left( \vect{X}_{1}^{n} ; Y_{1}^{n} \right)  
\end{equation}
where we use $\vect{X}_{j}^{k}$ to denote $\vect{X}_{j}, \ldots,
\vect{X}_{k}$ and 
where the supremum is over the set of all probability distributions on
$\vect{X}_{1}^{n}$ satisfying the constraints, \emph{i.e.},
\begin{equation}
  \label{eq:peak_constraint}
  \|\vect{X}_{k}\|^{2} \leq \Es, \quad \text{almost surely}, 
  \quad k = 1, 2, \ldots, n   
\end{equation}
for a peak constraint, or 
\begin{equation}
  \label{eq:average_constraint}
  \frac{1}{n} \sum_{k=1}^{n} \E{\|\vect{X}_{k}\|^{2}} \leq \Es   
\end{equation}
for an average constraint.

Specializing \cite[Theorem~4.2]{lapidothmoser03_3} or
\cite[Theorem~6.10]{moser04}, respectively, to MISO fading, we have
\begin{equation}
  \label{eq:villa}
  \varlimsup_{\SNR \uparrow \infty}
  \Big\{ \const{C}(\SNR) - \log \log \SNR  \Big\}
  < \infty.  
\end{equation}
The fading number $\chi$ is now defined as in
\cite[Definition~4.6]{lapidothmoser03_3} and in
\cite[Definition~6.13]{moser04} by
\begin{equation}
  \label{eq:def_general_chi_intro}
  \chi(\{\trans{\vect{H}}_k\}) \eqdef  \varlimsup_{\SNR
    \uparrow \infty}  
  \Big\{ \const{C}(\SNR) - \log \log \SNR
  \Big\}. 
\end{equation}
\emph{Prima facie} the fading number depends on whether a peak-power
constraint \eqref{eq:peak_constraint} or an average-power constraint
\eqref{eq:average_constraint} is imposed on the input. Since a
peak-power constraint is more stringent than an average-power
constraint, we will derive the upper bound using the average-power
constraint and the lower bound using the peak-power constraint. In
case of an isotropically distributed fading process we shall see that
both constraints lead to identical fading numbers.

\section{Main Results}
\label{sec:main}

We first state a new upper bound to the fading number of a MISO fading
channel: 
\begin{theorem}
  \label{thm:uppper-bound}
  Consider a MISO fading channel with memory \eqref{eq:channel_model}
  where the stationary and ergodic fading process $\{\vect{H}_k\}$
  takes value in $\Complex^{\nt}$ and satisfies $h(\{\vect{H}_k\}) >
  -\infty$ and $\E{\|\vect{H}_k\|^{2}} < \infty$.  Then, irrespective
  of whether a peak-power constraint \eqref{eq:peak_constraint} or an
  average-power constraint \eqref{eq:average_constraint} is imposed on
  the input, the fading number $\chi\big(\{\trans{\vect{H}}_k\}\big)$
  is upper-bounded by
  \begin{IEEEeqnarray}{rCl}
    \chi \big( \{\trans{\vect{H}}_{k}\} \big)  
    & \le & \sup_{\hvect{x}_{-\infty}^0} \Big\{ \log\pi + \E{\log
      |\trans{\vect{H}}_0 \hvect{x}_0|^2} \nonumber \\
    && \qquad\qquad-\:
    h\big(\trans{\vect{H}}_0\hvect{x}_0 \,\big|\, \{
    \trans{\vect{H}}_{\ell}\hvect{x}_{\ell} \}_{\ell=-\infty}^{-1} \big)
    \Big\}
    \label{eq:upperbound}    
  \end{IEEEeqnarray}
  where $\hvect{x}_{\ell} \eqdef
  \frac{\vect{x}_{\ell}}{\|\vect{x}_{\ell}\|}$ denotes a vector of
  unit length.
\end{theorem}
\begin{proof}
  The proof is in part pretty technical. We therefore give only an
  outline and omit the details.
  
  Similar to the derivation of the SIMO fading number with memory, the
  proof starts with a lemma that limits the possible joint input
  distributions on $\vect{X}_1, \ldots, \vect{X}_n$ to such under
  which each random vector $\vect{X}_{\ell}$ has the same law with an
  average power equal to the constraint $\Es$. Unfortunately, the
  proof is complicated by the fact that this lemma does not guarantee
  equal marginals for the time epochs $k$ on the border of a block.
  However, these edge effects wash out once we let the blocklength $n$
  tend to infinity.
  
  The proof then proceeds as follows: the mutual information between
  joint input and joint output is split up into a term describing the
  memoryless case and a term that takes care of the memory:
  \begin{IEEEeqnarray}{rCl}
    \IEEEeqnarraymulticol{3}{l}{
      \lim_{n\to\infty} \frac{1}{n} I(\vect{X}_1^n;
      Y_1^n) }\nonumber\\\qquad
    & \le & \lim_{n\to\infty} \frac{1}{n} \sum_{k=1}^n 
    \Big( I(\vect{X}_k; Y_k) \nonumber \\
    && \qquad\qquad +\: I\big(\trans{\vect{H}_k}\hvect{X}_k;
    \{\trans{\vect{H}_{\ell}}\hvect{X}_{\ell}\}_{\ell=1}^{k-1} \big|
    \hvect{X}_1^k \big) \Big),    
  \end{IEEEeqnarray}  
  where the above mentioned lemma guarantees an input distribution
  with equal marginals and an average power of $\Es$. 
  
  The first term is then upper-bounded by 
  \begin{equation}
    I(\vect{X}_k;  Y_k) \lessapprox \E[\hvect{X}_k]{
      \const{C}_{\textnormal{SISO,IID,} 
        H=\trans{\vect{H}_k}\hvect{X}_k}(\Es)}
  \end{equation}
  where the approximation results from ignoring some additional terms
  that tend to zero as $\Es$ tends to infinity. We hence get a bound
  \begin{IEEEeqnarray}{rCl}
    \const{C} & \lessapprox &  \Exp_{\hvect{X}_{-\infty}^0} \Big[
    \const{C}_{\textnormal{SISO,IID,} 
      H=\trans{\vect{H}_0}\hvect{X}_0}(\Es) \nonumber \\
    && \quad +\: I\big(
    \trans{\vect{H}}_0\hvect{X}_0 ;  
    \{\trans{\vect{H}}_{\ell}\hvect{X}_{\ell}\}_{\ell=-\infty}^{-1} 
    \,\big|\,
    \{\hvect{X}_{\ell}=\hvect{x}_{\ell}\}_{\ell=-\infty}^{0}\big) \Big]
    \\
    & \le & \sup_{\hvect{x}_{-\infty}^0} \left\{
      \const{C}_{\textnormal{SISO,IID,} 
        H=\trans{\vect{H}_0}\hvect{x}_0}(\Es) + I\big(
      \trans{\vect{H}}_0\hvect{x}_0 ;  
      \{\trans{\vect{H}}_{\ell}\hvect{x}_{\ell}\}_{\ell=-\infty}^{-1} 
    \right\}
    \nonumber\\
  \end{IEEEeqnarray} 
  where the approximation results from ignoring the edge effects and
  the terms that will tend to zero as $\Es$ tends to infinity.

  The claim now follows by using the fading number of a memoryless
  SISO fading channel.
\end{proof}

Next we state a lower bound to the fading number of a MISO fading
channel:
\begin{theorem}
  \label{thm:lowerbound}
  Consider a MISO fading channel with memory \eqref{eq:channel_model}
  where the stationary and ergodic fading process $\{\vect{H}_k\}$
  takes value in $\Complex^{\nt}$ and satisfies $h(\{\vect{H}_k\}) >
  -\infty$ and $\E{\|\vect{H}_k\|^{2}} < \infty$.  Then the fading
  number $\chi\big(\{\trans{\vect{H}}_k\}\big)$ is lower-bounded by
  \begin{IEEEeqnarray}{rCl}
    \chi \big( \{\trans{\vect{H}}_{k}\} \big)  
    & \ge & \sup_{\hvect{x}} \Big\{ \log\pi + \E{\log
      |\trans{\vect{H}}_0 \hvect{x}|^2} \nonumber \\
    && \qquad\qquad-\:
    h\big(\trans{\vect{H}}_0\hvect{x} \,\big|\, \{ 
    \trans{\vect{H}}_{\ell}\hvect{x} \}_{\ell=-\infty}^{-1} \big)
    \Big\} 
    \label{eq:lowerbound}    
  \end{IEEEeqnarray}
  where $\hvect{x} \eqdef
  \frac{\vect{x}}{\|\vect{x}\|}$ denotes a vector of
  unit length. 

  Moreover, this lower bound is achievable by IID inputs that can be
  expressed as the product of a constant unit vector
  $\hvect{x}\in\Complex^{\nt}$ and a circularly symmetric, scalar,
  complex IID 
  random process $\{X_k\}$ such  that 
  \begin{equation}
    \log |X_k|^2 \sim \Uniform{[\log\log\Es, \log\Es]}.
  \end{equation}
  Note that this input satisfies the peak-power constraint
  \eqref{eq:peak_constraint} (and therefore also the average-power
  constraint \eqref{eq:average_constraint}). 
\end{theorem}
\begin{proof}
  We only give an outline of the proof. The details are omitted.

  The lower bound is based on the assumption of a specific input
  distribution which is chosen to be of the form
  \begin{equation}
    \vect{X}_k = X_k \cdot \hvect{x}
  \end{equation}
  where $\hvect{x}$ is a deterministic unit vector (the
  beam-direction) and where $\{X_k\}$ is IID circularly symmetric with
  \begin{equation}
    \log|X_k|^2 \sim \Uniform{[\log \log\Es, \log\Es]}.
  \end{equation}
  Note that this choice for $\{X_k\}$ achieves the fading number for
  the SISO fading channel
  \begin{equation}
    Y_k = (\trans{\vect{H}_k}\hvect{x})\cdot X_k + Z_k
  \end{equation}
  with fading process $\{H_k\} = \{\trans{\vect{H}_k}\hvect{x}\}$.
  The lower bound is then derived by proving
  \begin{equation}
    \frac{1}{n} I(\vect{X}_1^n; Y_1^n) \approx \frac{1}{n} \sum_{k=1}^n
    I\big( X_k; Y_k \big|
    \{\trans{\vect{H}_{\ell}}\hvect{x}\}_{\ell=1}^{k-1} \big)
  \end{equation}
  and using the results of memoryless SISO fading channels with
  side-information \cite{lapidothmoser03_3}, \cite{moser04}. 
\end{proof}

\section{Special Case of Isotropically Distributed Fading}
\label{sec:csfading}

We next consider the special case of isotropically distributed fading
processes, \emph{i.e.}, for every deterministic unitary $\nt\times\nt$
matrix $\mat{U}$
\begin{equation}
  \label{eq:eqlaw}
  \vect{H}_k \eqlaw \mat{U}\vect{H}_k,
\end{equation}
where we use ``$\eqlaw$'' to denote \emph{equal in law}.

In this case we have the following corollary:
\begin{corollary}
  \label{cor:cs}
  Consider a MISO fading channel with memory \eqref{eq:channel_model}
  where the stationary and ergodic fading process $\{\vect{H}_k\}$
  takes value in $\Complex^{\nt}$, satisfies $h(\{\vect{H}_k\}) >
  -\infty$ and $\E{\|\vect{H}_k\|^{2}} < \infty$, and is
  \emph{isotropically distributed}.  Then the upper bound
  \eqref{eq:upperbound} and the lower bound \eqref{eq:lowerbound}
  coincide and the fading number
  $\chi_{\textnormal{iso}}\big(\{\trans{\vect{H}}_k\}\big)$ is given by
  \begin{IEEEeqnarray}{rCl}
    \chi_{\textnormal{iso}} \big( \{\trans{\vect{H}}_{k}\} \big)  
    & = & \log\pi + \E{\log
      |\trans{\vect{H}}_0 \hvect{e}|^2} \nonumber \\
    && -\:
    h\big(\trans{\vect{H}}_0\hvect{e} \,\big|\, \{ 
    \trans{\vect{H}}_{\ell}\hvect{e} \}_{\ell=-\infty}^{-1} \big)
    \label{eq:csfadingnumber}    
  \end{IEEEeqnarray}
  where $\hvect{e}$ is any deterministic unit vector.
\end{corollary}
\begin{proof}
  This follows immediately from Theorem~\ref{thm:uppper-bound} and
  \ref{thm:lowerbound} by noting that for any $\hvect{e}$
  \begin{equation}
    \trans{\vect{H}}_k\hvect{e} \eqlaw
    \trans{\vect{H}}_k\trans{\mat{U}}\hvect{e}  = 
    \trans{\vect{H}}_k\hvect{e}'
  \end{equation}
  where the first equality in law follows from \eqref{eq:eqlaw} and
  the second equality by defining a new unit vector $\hvect{e}'\eqdef
  \trans{\mat{U}}\hvect{e}$. Note that for the MISO case
  \emph{isotropically distributed} is equivalent to \emph{rotation
    commutative in the generalized sense} as defined in
  \cite[Defintion~4.37]{lapidothmoser03_3} or
  \cite[Defintion~6.37]{moser04}.
\end{proof}

\section{Special Case of Gaussian Fading}
\label{sec:gaussianfading}

In this section we assume that the fading process $\{\vect{H}_k\}$ is
a mean-$\vect{d}$ Gaussian process such that
$\{\tilde{\vect{H}}_k\}=\{\vect{H}_k -\vect{d}\}$ is a zero-mean,
circularly symmetric, stationary, ergodic, complex Gaussian process
with matrix-valued spectral distribution function $\mat{F}(\cdot)$,
and with covariance matrix $\mat{K}$. Furthermore, we assume that the
prediction error covariance matrix $\mat{\Sigma}$ is non-singular
(regularity assumption).

\subsection{Upper Bound for Gaussian Fading}
\label{sec:gaussupper}

We start with a new derivation of the upper bound \eqref{eq:gauss}
based on Theorem~\ref{thm:uppper-bound}. We will see that
\eqref{eq:gauss} is in general less tight than \eqref{eq:upperbound}.

We start by loosening the upper bound \eqref{eq:upperbound} as
follows:
\begin{IEEEeqnarray}{rCl}
  \IEEEeqnarraymulticol{3}{l}{
    \chi \big( \{\trans{\vect{H}}_{k}\} \big) }\nonumber\\\quad
  & \le &
  \sup_{\hvect{x}_{-\infty}^0} \Big\{ \log\pi + \E{\log
    |\trans{\vect{H}}_0 \hvect{x}_0|^2} \nonumber \\
  && \qquad\quad -\:
  h\big(\trans{\vect{H}}_0\hvect{x}_0 \,\big|\, \{
  \trans{\vect{H}}_{\ell}\hvect{x}_{\ell} \}_{\ell=-\infty}^{-1} \big)
  \Big\}
  \\
  & = & \sup_{\hvect{x}_{-\infty}^0} \Big\{ \log\pi + \E{\log
    |\trans{\vect{H}}_0 \hvect{x}_0|^2} -
  h\big(\trans{\vect{H}}_0\hvect{x}_0\big)
  \nonumber \\
  && \qquad\quad +\: h\big(\trans{\vect{H}}_0\hvect{x}_0\big) -
  h\big(\trans{\vect{H}}_0\hvect{x}_0 \,\big|\, \{
  \trans{\vect{H}}_{\ell}\hvect{x}_{\ell} \}_{\ell=-\infty}^{-1} \big)
  \Big\}
  \\
  & \le & \sup_{\hvect{x}_0} \Big\{ \log\pi + \E{\log
    |\trans{\vect{H}}_0 \hvect{x}_0|^2} -
  h\big(\trans{\vect{H}}_0\hvect{x}_0\big) \Big\}
  \nonumber \\
  && +\: \sup_{\hvect{x}_{-\infty}^0} \Big\{
  h\big(\trans{\vect{H}}_0\hvect{x}_0\big) -
  h\big(\trans{\vect{H}}_0\hvect{x}_0 \,\big|\, \{
  \trans{\vect{H}}_{\ell}\hvect{x}_{\ell} \}_{\ell=-\infty}^{-1} \big)
  \Big\}
  \IEEEeqnarraynumspace
  \label{eq:gnurk22}
  \\
  & = & \sup_{\hvect{x}_0} \Big\{ \log\pi + \E{\log
    |\trans{\vect{H}}_0 \hvect{x}_0|^2} -
  h\big(\trans{\vect{H}}_0\hvect{x}_0\big) \Big\}
  \nonumber \\
  && +\: \sup_{\hvect{x}_{-\infty}^0}
  I\big(\trans{\vect{H}}_0\hvect{x}_0 ; \{
  \trans{\vect{H}}_{\ell}\hvect{x}_{\ell} \}_{\ell=-\infty}^{-1}
  \big),
  \label{eq:upperbound2}
\end{IEEEeqnarray}
where \eqref{eq:gnurk22} follows from
\begin{equation}
  \sup_x \{ f(x) + g(x)\} \le \sup_x f(x) + \sup_x g(x).
\end{equation}

In \cite[Corollary~4.28]{lapidothmoser03_3},
\cite[Corollary~6.28]{moser04} it has been shown that the IID MISO
fading number \eqref{eq:iidmiso} for Gaussian fading is given by
\begin{IEEEeqnarray}{rCl}
  \chi(\trans{\vect{H}}) & = & \sup_{\|\hvect{x}\|=1} \left\{ \log \pi +
    \E{\log|\trans{\vect{H}} \hvect{x}|^2} -
    h\big(\trans{\vect{H}}\hvect{x}\big) \right\} 
  \\
  & = & -1 + \log d_{\ast}^2 -
  \Ei{-d_{\ast}^2} 
  \label{eq:iidmisogauss}  
\end{IEEEeqnarray}
where $d_{\ast}$ is given in \eqref{eq:dgauss}. This proves the
equivalence of the first supremum in \eqref{eq:upperbound2} with the
first three terms of \eqref{eq:gauss}. It therefore only remains to
prove that
\begin{equation}
  \sup_{\hvect{x}_{-\infty}^0} I\big(\trans{\vect{H}}_0\hvect{x}_0 ; \{
    \trans{\vect{H}}_{\ell}\hvect{x}_{\ell} \}_{\ell=-\infty}^{-1} \big)
    \le \log\frac{\|\mat{K}\|}{\lambda_{\textnormal{min}}}.
    \label{eq:derivgaussupper0}
\end{equation}
To this goal note that
\begin{IEEEeqnarray}{rCl}
  \IEEEeqnarraymulticol{3}{l}{
    \sup_{\hvect{x}_{-\infty}^0} I\big(\trans{\vect{H}}_0\hvect{x}_0 ; \{
    \trans{\vect{H}}_{\ell}\hvect{x}_{\ell} \}_{\ell=-\infty}^{-1} \big)
    }\nonumber\\\quad
    & \le & 
    \sup_{\hvect{x}_{-\infty}^0} I\big(\trans{\vect{H}}_0\hvect{x}_0 ; \{
    \trans{\vect{H}}_{\ell}\hvect{x}_{\ell} \}_{\ell=-\infty}^{-1},
    \vect{H}_{-\infty}^{-1} \big) 
    \\
    & = & 
    \sup_{\hvect{x}_0} I\big(\trans{\vect{H}}_0\hvect{x}_0 ; 
    \vect{H}_{-\infty}^{-1} \big) 
    \\    
    & = & 
    \sup_{\hvect{x}_0} \Big\{ h\big(\trans{\vect{H}}_0\hvect{x}_0\big)
    -  h\big(\trans{\vect{H}}_0\hvect{x}_0\,\big|\,
    \vect{H}_{-\infty}^{-1} \big) \Big\}
    \\    
    & = & 
    \sup_{\hvect{x}_0} \left\{ \log \left(\pi e \hermi{\hvect{x}_0}
    \mat{K} \hvect{x}_0 \right)
    -  h\big(\trans{\vect{H}}_0\hvect{x}_0\,\big|\,
    \vect{H}_{-\infty}^{-1} \big) \right\}.
    \label{eq:derivgaussupper1}
\end{IEEEeqnarray}
Here, the first inequality follows from the inclusion of additional
random variables in the mutual information; the subsequent equality
from the fact that given the past realization of the fading,
$\trans{\vect{H}}_{0}\hvect{x}_{0}$ is independent of $\{
\trans{\vect{H}}_{\ell}\hvect{x}_{\ell} \}_{\ell=-\infty}^{-1}$; and
in the last equality we have used the expression for the differential
entropy of a Gaussian random variable with $\mat{K}$ denoting the
covariance matrix of $\{\vect{H}_k\}$.

Note that the first inequality in general is not tight, \emph{i.e.},
\eqref{eq:gauss} is in general looser than \eqref{eq:upperbound2}
which in turn is in general looser than \eqref{eq:upperbound}.

To compute the second term on the RHS of \eqref{eq:derivgaussupper1},
we express the fading $\vect{H}_0$ as
\begin{equation}
  \vect{H}_0 = \bar{\vect{H}}_0 + \tilde{\vect{H}}_0
\end{equation}
with $\bar{\vect{H}}_0$ being the best estimate of $\vect{H}_0$ based
on the past realizations. We note that
$\tilde{\vect{H}}_0\sim\NormalC{\vect{0}}{\mat{\Sigma}}$  where
$\mat{\Sigma}$ denotes the prediction error covariance matrix. Hence
\begin{equation}
  h\big(\trans{\vect{H}}_0\hvect{x}_0\,\big|\,
    \vect{H}_{-\infty}^{-1} \big) = \log \left(\pi e \hermi{\hvect{x}_0}
      \mat{\Sigma}\hvect{x}_0\right). 
\end{equation}
The bound \eqref{eq:derivgaussupper0} now follows by the Rayleigh-Ritz
Theorem \cite[Theorem~4.2.2]{hornjohnson85},
\cite[Theorem~A.9]{moser04}
\begin{equation}
  \lambda_{\textnormal{min}} = \min_{\hvect{x}} \hermi{\hvect{x}}
  \mat{\Sigma}\hvect{x},
\end{equation}
the definition of the Euclidean norm of matrices, and the
properties of positive semi-definite matrices:
\begin{equation}
  \max_{\hvect{x}} \hermi{\hvect{x}} \mat{K}\hvect{x} = 
  \max_{\hvect{x}} \hermi{\hvect{x}} \trans{\mat{S}}\mat{S}\hvect{x} = 
  \max_{\hvect{x}} \|\mat{S}\hvect{x}\|^2 = \|\mat{S}\|^2 =
  \|\mat{K}\|.  
\end{equation}

\subsection{Spatially IID Gaussian Fading}
\label{sec:tild-being-spat}

We next specialize the assumptions to the case where
$\{\tilde{\vect{H}}_k\}=\{\vect{H}_k -\vect{d}\}$ is a spatially IID
process where each component is a zero-mean unit-variance circularly
symmetric complex Gaussian process of spectral distribution function
$\const{F}(\cdot)$. For this case we will now present a new derivation
of the result \eqref{eq:iidgauss} based on our new bounds.

Note that we cannot apply Corollary~\ref{cor:cs} here: even though
$\{\tilde{\vect{H}}_k\}$ is isotropically distributed,
$\{\vect{H}_k\}$ is not due to its mean vector $\vect{d}$.

However, the term $I\big(\trans{\vect{H}}_0\hvect{x}_0 ; \{ 
\trans{\vect{H}}_{\ell}\hvect{x}_{\ell} \}_{\ell=-\infty}^{-1}
\big)$ does not depend on the particular choice of
$\hvect{x}_{\ell}$:
\begin{IEEEeqnarray}{rCl}
  \IEEEeqnarraymulticol{3}{l}{
    I\big(\trans{\vect{H}}_0\hvect{x}_0 ; \{ 
  \trans{\vect{H}}_{\ell}\hvect{x}_{\ell} \}_{\ell=-\infty}^{-1}
  \big)
  }\nonumber\\\quad
  & = & I\big(\trans{\vect{H}}_0\hvect{x}_0 -
  \trans{\vect{d}}\hvect{x}_0 ; \{ 
  \trans{\vect{H}}_{\ell}\hvect{x}_{\ell} -
  \trans{\vect{d}}\hvect{x}_{\ell}\}_{\ell=-\infty}^{-1} 
  \big)
  \\
  & = & I\big(\trans{\tilde{\vect{H}}}_0\hvect{x}_0 ; \{ 
  \trans{\tilde{\vect{H}}}_{\ell}\hvect{x}_{\ell} \}_{\ell=-\infty}^{-1}
  \big)
  \\
  & = & I\big(\trans{\tilde{\vect{H}}}_0\hvect{e} ; \{ 
  \trans{\tilde{\vect{H}}}_{\ell}\hvect{e} \}_{\ell=-\infty}^{-1}
  \big)
  \\
  & = & I \big( H_0^{(1)} ; \{H_{\ell}^{(1)}\}_{\ell=-\infty}^{-1}
  \big) 
  \\
  & = & \log \frac{1}{\eps^2}.
\end{IEEEeqnarray}
Equation~\eqref{eq:iidgauss} now follows from
\eqref{eq:iidmisogauss}, Theorem~\ref{thm:uppper-bound}, and
Theorem~\ref{thm:lowerbound} by noting that
\begin{equation}
  \max_{\|\hvect{x}\|=1} \frac{\left|
      \E{\trans{\vect{H}}_k}\hvect{x}
    \right|}{\sqrt{\Var{\trans{\vect{H}_k}\hvect{x}}}} =
  \max_{\|\hvect{x}\|=1} \left|  \trans{\vect{d}}\hvect{x}
  \right|= \|\vect{d}\|,
\end{equation}
where the maximum is achieved for $\hvect{x}=\vect{d}/\|\vect{d}\|$.

\section{Discussion \& Conclusion}
\label{sec:conclusion}

We have derived two bounds for a MISO fading channel of general law
including memory. Both bounds show the same structure involving the
maximization of a deterministic beam-direction $\hvect{x}$, which
suggests that beam-forming is optimal at high $\SNR$. However, one has
to be aware that the beam-direction is not chosen to maximize the
$\SNR$, but to maximize the fading number.

The differences between the upper and lower bound lies in the details
of the maximization: while in the lower bound one single direction
unit vector $\hvect{x}$ is chosen for all time, the upper bound allows
for different realizations of $\hvect{x}_k$ for different times $k$.

We are convinced that the lower bound is actually tight: intuition
tells that for our stationary channel model a stationary input should
be sufficient for achieving the capacity. As a matter of fact in the
SISO and SIMO case it has been shown that actually an IID input
suffices to achieve capacity at high $\SNR$ \cite{lapidothmoser03_3},
\cite{moser04}, \cite{lapidothmoser04_3app}. Furthermore, we have been
able to modify the derivation of the upper bound such as to get the
following bound:
\begin{IEEEeqnarray}{rCl}
  \IEEEeqnarraymulticol{3}{l}{
    \chi(\{\trans{\vect{H}_k}\}) }\nonumber\\\quad
  & \le &\sup_{Q_{\hvect{X}_{-\kappa},
      \ldots, \hvect{X}_0}\in\set{P}^*}  
  \Exp_{\hvect{X}_{-\kappa}^{0}} \bigg[ \log\pi +
  \Econd{\trans{\vect{H}_0}\hvect{X}_0}{\hvect{X}_0}
  \nonumber \\
  && \qquad -\: h\big(\trans{\vect{H}}_0\hvect{X}_0 \;\big|\;
  \{\trans{\vect{H}}_{\ell}\hvect{X}_{\ell}\}_{\ell=-\kappa}^{-1},
  \{\hvect{X}_{\ell}=\hvect{x}_{\ell}\}_{\ell=-\kappa}^{0}\big)
  \bigg]. 
  \IEEEeqnarraynumspace
\end{IEEEeqnarray}
Here, $\kappa$ is an arbitrary (large) positive integer, and
$\set{P}^*$ denotes the set of all joint distributions on
$\hvect{X}_{-\kappa}, \ldots, \hvect{X}_0$ where
\begin{equation}
  \|\hvect{X}_{\ell}\| = 1 \qquad \textnormal{a.s.}
\end{equation}
and where each $\hvect{X}_{\ell}$ has the same distribution, for all
$\ell=-\kappa, \ldots, 0$.  This bound can easily be loosened to get
the result of Theorem~\ref{thm:uppper-bound}: one simply upper-bounds
the supremum over $Q$ with a supremum over $\hvect{x}_{-\kappa}^0$.
However, it intuitively seems that under the constraint that all
marginals of $Q$ must be the same this bound is not tight.

In the case of isotropically distributed fading the particular choice
of direction has no influence on the fading process and therefore the
upper and lower bounds coincide.

In the case of Gaussian fading we could show that the bounds presented
in \cite{kochlapidoth05_1} and \cite{kochlapidoth05_2} are special
cases of the new bounds presented here, where the new upper bound
\eqref{eq:upperbound} is in general tighter than \eqref{eq:gauss}.

The success of further attempts on deriving the MISO fading number
precisely will be crucial to the investigation of the fading number
of general MIMO fading channels.

\section*{Acknowledgments}

Helpful comments from Amos Lapidoth, Tobias Koch, and Daniel H\"osli
are gratefully acknowledged.

\bibliographystyle{IEEEtran}
\bibliography{header,bibliofile}

\end{document}